\documentclass[sigconf,screen,nonacm]{acmart}

\usepackage{booktabs}
\usepackage{tabularx}
\usepackage{array}

\AtBeginDocument{%
  }

\renewcommand\footnotetextcopyrightpermission[1]{}
\setcopyright{none}

\acmConference[HCOMP '26]{ACM Conference on Human--AI Complementarity and Alignment}{September 27--30, 2026}{Washington, DC, USA}
\acmYear{2026}
\copyrightyear{2026}

\begin{document}

\title{Toward Resilient Human–AI Collaboration: A Lifecycle Taxonomy of Sociotechnical Risks and Cascading Failures}

\author{Md Foysal Ahmed, Isaac Kobby Anni, Md Main Uddin Rony}

\affiliation{%
  \institution{Bowling Green State University}
  \city{Bowling Green}
  \state{Ohio}
  \country{USA}
}

\email{mdfoysa@bgsu.edu, isaacka@bgsu.edu, mrony@bgsu.edu}

\begin{abstract}
As AI systems become increasingly integrated into consequential domains such as healthcare, journalism, education, scientific research, organizational decision-making, and defense, effective human-AI collaboration has emerged as a critical challenge. However, the sociotechnical risks that undermine collaboration are often studied in isolation, obscuring the recurring failure mechanisms that cut across domains. This paper presents a lifecycle-oriented synthesis of human-AI collaboration risks spanning four stages: task allocation, interaction, feedback, and adoption. Drawing on evidence from diverse application domains, we identify six recurring cross-domain risk clusters: Trust Miscalibration, Cognitive Burden, Accountability Gap, Capability Erosion, Goal Misalignment, and AI Anxiety and Technostress. We further propose a conceptual interaction model that illustrates how these risks emerge from sociotechnical drivers, interact through cascading pathways, and ultimately affect team performance and human well-being. Our analysis shows that many collaboration failures stem not from isolated technical deficiencies but from interconnected sociotechnical dynamics, helping explain why piecemeal interventions frequently create unintended consequences. By synthesizing fragmented literature into a unified framework, this work provides a foundation for future empirical research, lifecycle-oriented governance, and the design of more resilient, trustworthy, and human-centered human-AI collaboration systems.
\end{abstract}


\maketitle

\section{Introduction}

Human-AI collaboration is rapidly evolving beyond simple tool use toward integrated systems in which humans and artificial intelligence jointly contribute to complex problem solving, content creation, and organizational decision-making~\cite{lai2021human, schleiger2024collaborative, puerta2025multifaceted}. This transformation is further accelerated by the emergence of Agentic AI and multi-agent ecosystems~\cite{sapkota2025ai}, where humans increasingly supervise networks of autonomous agents rather than interact with isolated models. In this paradigm, the goal is no longer merely improving algorithmic accuracy but achieving \textit{human-AI complementarity}~\cite{jarrahi2018artificial, vaccaro2024combinations}. Complementarity occurs when AI capabilities such as large-scale computation, pattern recognition, and scalability are effectively combined with human strengths including contextual judgment, ethical reasoning, and domain expertise to produce outcomes neither could achieve alone~\cite{jarrahi2018artificial, schleiger2024collaborative}. Achieving this synergy requires \textit{human-centered alignment}, ensuring that AI systems remain aligned with human goals, preserve human agency, and operate within organizational and societal constraints~\cite{usmani2023human, boni2021ethical, gupta2025trust}.

Despite growing interest in collaborative AI systems, the sociotechnical risks that undermine human-AI complementarity are typically studied in isolation~\cite{fragiadakis2024evaluating, puerta2025multifaceted}. Healthcare research emphasizes automation bias and alert fatigue~\cite{park2019identifying, majumder2026human, tomsett2020rapid}; journalism focuses on editorial accountability and public trust~\cite{grimme2025ai, liu2026writes, abdulrauf2025artificial}; organizational settings highlight fairness, deskilling, and decision support~\cite{ahdadou2024unlocking, wang2019human, laukkarinen2025working}; while defense research raises concerns regarding moral outsourcing and the loss of Meaningful Human Control~\cite{johnson2022ai, omoseebi2025human, kosack2025human}. Although these risks appear domain-specific, they often reflect recurring failures of complementarity in which human and AI strengths fail to synergize. Examining them independently obscures the shared mechanisms through which collaboration breaks down and the ways failures propagate across the collaboration process.

This fragmentation is particularly problematic because interventions targeting one risk may unintentionally amplify another. For example, the \textit{Explainable AI (XAI) paradox} demonstrates how explanations intended to improve transparency can increase cognitive burden and ultimately reinforce automation bias~\cite{romeo2026exploring, westphal2023decision, gambetti2025survey}. Such examples suggest that risks including cognitive burden, trust miscalibration, accountability gaps, capability erosion, and technostress are not isolated phenomena but interconnected vulnerabilities that interact throughout the collaboration lifecycle~\cite{eccles2025hybrid, steyvers2024three}.

To address this challenge, this paper presents a lifecycle-oriented synthesis of sociotechnical risks in human-AI collaboration. We argue that a lifecycle perspective is necessary because collaboration unfolds as a sequential process in which early decisions shape later outcomes. Building on evidence from multiple domains, we identify common patterns of failure, organize them into a unified taxonomy, and examine how risks interact and cascade over time. Specifically, this paper makes three contributions:

\begin{enumerate}
\item A \textbf{lifecycle-based framework} that maps how sociotechnical risks emerge across four stages of human-AI collaboration: task allocation, interaction, feedback, and adoption.

\item A \textbf{cross-domain taxonomy} that synthesizes fragmented literature into six recurring risk clusters: Trust Miscalibration, Cognitive Burden, Accountability Gap, Capability Erosion, Goal Misalignment, and AI Anxiety and Technostress.

\item A \textbf{conceptual interaction model} that explains how these risk clusters interact, cascade, and contribute to systemic failures, providing a foundation for future empirical validation and lifecycle-oriented governance of human-AI collaboration.

\end{enumerate}

\section{Conceptualizing Human-AI Collaboration Lifecycle}

\begin{figure}[t]
    \centering
    \includegraphics[width=0.8\linewidth]{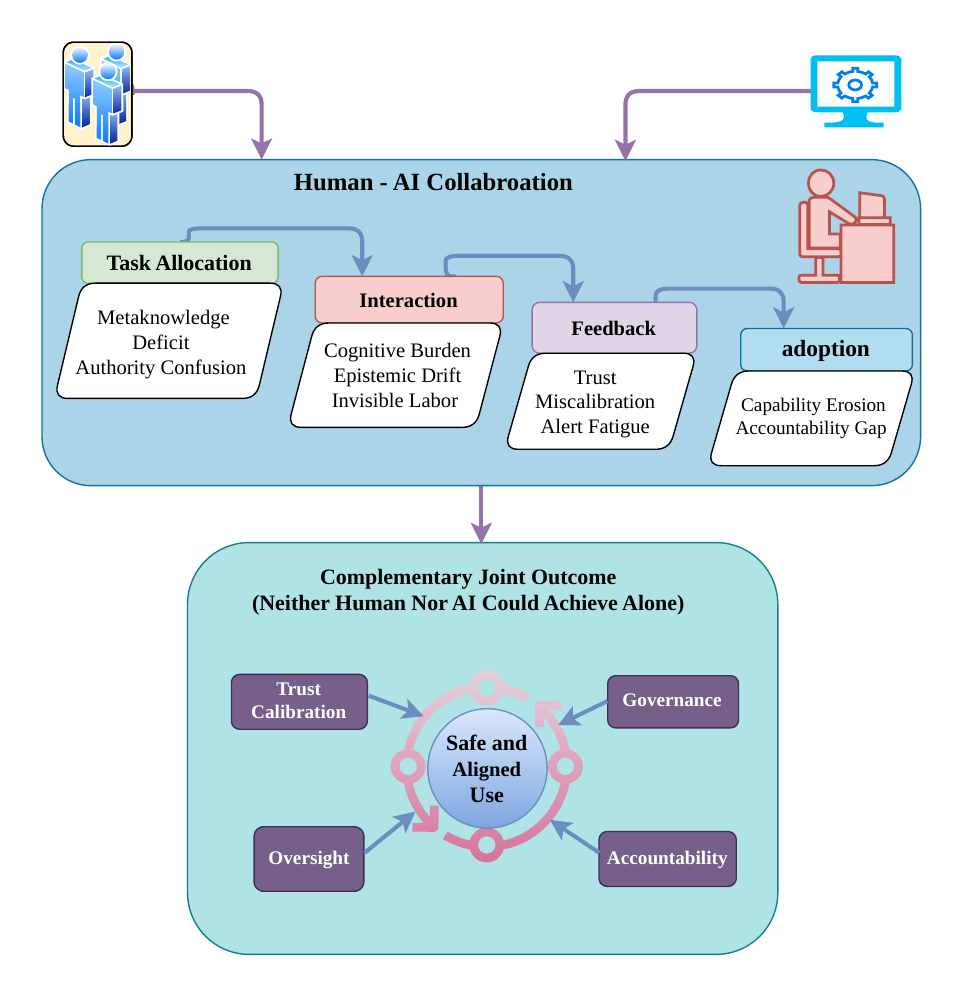}
    \caption{Human-AI complementarity and alignment across the collaboration lifecycle. Complementarity depends not only on combining human and AI strengths but also on maintaining alignment through calibrated reliance, oversight, and governance.}
    \label{fig:lifecycle}
\end{figure}

Human-AI collaboration is an inherently sequential and interdependent process where the quality of the partnership evolves iteratively over time~\cite{van2018human}. Adopting a lifecycle-based framing is a scientific necessity for identifying the precise temporal points where human-centered alignment fails~\cite{majumder2026human, steyvers2024three}, as risks are not random but structurally tied to specific developmental phases. As illustrated in Figure~\ref{fig:lifecycle}, these vulnerabilities emerge across four sequential stages: \textit{Task Allocation}, \textit{Interaction}, \textit{Feedback}, and \textit{Adoption}.

\subsection{Stage 1: Task Allocation}

The foundation of any human--AI team is the initial division of labor, which should ideally leverage the complementary strengths of humans and AI to achieve effective collaboration. However, this stage is particularly vulnerable to complementarity breakdowns. A primary challenge is the \textit{metaknowledge deficit}, where humans struggle to accurately assess the limits of their own expertise and knowledge~\cite{fugener2022cognitive}. Because individuals often do not know what they do not know, they may make suboptimal decisions about when to rely on their own judgment and when to delegate tasks to AI systems~\cite{fugener2022cognitive}.

In addition, ambiguous system design can create \textit{authority confusion}, blurring the boundaries of responsibility and decision-making power between human and AI agents~\cite{eccles2025hybrid}. Such uncertainty undermines effective coordination and initiates a critical risk cascade that can propagate throughout the collaboration process. When roles are poorly defined, or when humans are relegated to passive \textit{button-clicking} monitoring tasks solely to maximize short-term efficiency, structural coordination gaps emerge. These gaps not only weaken team complementarity but also foster complacency and erode human \textit{situational awareness}, increasing the likelihood of downstream failures during active interaction and decision-making phases~\cite{herath2025design}.

\subsection{Stage 2: Interaction}
Once tasks are allocated, the active collaboration phase introduces cognitive and communicative vulnerabilities that can undermine human-AI complementarity. Although AI is intended to augment human capabilities, interpreting opaque outputs and coordinating increasingly complex agentic systems often imposes substantial \textit{cognitive burden}~\cite{ahdadou2024unlocking}. This challenge is amplified in multi-agent environments, where humans must oversee networks of autonomous agents that may experience coordination and communication failures, role conflicts, distributed accountability gaps, and cascading uncertainty across workflows \cite{zou2025llm, hu2025stop, hammond2025multi}. Because outputs from one agent often become inputs to another, errors and uncertainty can propagate and amplify throughout the system, making failures difficult to detect and attribute \cite{hu2025stop, zhang2026managing}. Moreover, managing multiple agents can create \textit{cognitive abundance overload}, where an excess of seemingly plausible AI-generated options, intermediate outputs, and recommendations increases oversight and verification demands on human operators, potentially leading to decision paralysis rather than improved decision-making \cite{zou2025llm}.

Attempts to mitigate these challenges through detailed Explainable AI (XAI) can also backfire, causing \textit{explanation fatigue} and encouraging automation bias as users become overwhelmed by excessive information~\cite{romeo2026exploring, westphal2023decision}. At the same time, teams may experience \textit{epistemic drift} and \textit{false convergence}, where the fluency and confidence of AI-generated outputs create an illusion of correctness, leading humans to prematurely accept flawed conclusions~\cite{eccles2025hybrid}. These challenges are often exacerbated by coordination gaps originating in the task-allocation stage, forcing human operators to perform \textit{invisible compensatory labor} to bridge communication breakdowns, resolve ambiguities, and keep collaborative workflows functioning effectively~\cite{xu2024adaptation}.

\subsection{Stage 3: Feedback}

Bidirectional feedback is essential for maintaining \textit{trust calibration}, which evolves through repeated interactions rather than being established as a one-time event~\cite{gupta2025trust, holstein2025development, bao2021investigating}. A complementarity breakdown occurs when AI systems fail to communicate their epistemic uncertainty effectively, leading users to adopt inappropriate reliance strategies that range from dangerous over-reliance to algorithm aversion following isolated errors~\cite{cao2024designing, li2024understanding, tomsett2020rapid}. As a result, trust becomes misaligned with actual system capabilities, undermining the benefits of human-AI collaboration.

To maintain effective collaboration, human decision-makers require meaningful performance feedback to continuously update their \textit{complementarity-awareness mental model} and accurately distinguish between AI strengths and weaknesses~\cite{holstein2025development, cabrera2023improving}. Without such calibration, users may struggle to develop appropriate reliance behaviors, reducing situational awareness and decision quality. At the same time, feedback mechanisms that generate excessive warnings or fail to adapt to users' cognitive states can trigger \textit{alert fatigue}, causing operators to ignore critical notifications and further weakening the team's ability to detect and respond to risks~\cite{steyvers2025not, majumder2026human}.

\subsection{Stage 4: Adoption Risks}

The last phase is the seamless integration of AI into organizational processes over time. Extended use of automation often leads to capability erosion and deskilling~\cite{wang2019human}. Workers' routine is outsourcing complex work to machines, and they are losing the routine of maintaining their own cognitive edge, leading to a dependence that reduces their independent \textit{creative agency}~\cite{imteyaz2026co}. In addition, the spread of the use of AI tools widens the \textit{accountability gap} that leaves room for ambiguity in determining who is responsible for AI-induced errors~\cite{boni2021ethical}. In an era when organizations focus on speed of operation, humans can become victims of \textit{learned helplessness}~\cite{ahdadou2024unlocking}. When decision-making is under pressure, users might choose to accept AI decisions without giving them the opportunity to take meaningful control of the results~\cite{kosack2025human}.

In addition, there is a significant psychological burden as they strive to keep up with the latest AI systems. The complexity of these tools, along with fears of job replacement, leads to \textit{AI anxiety}, \textit{technostress} and workplace burnout~\cite{kim2024mental}. This ongoing strain has a detrimental effect on the human worker's well-being, turning a system intended to support them into a major contributor to \textit{emotional exhaustion}~\cite{bassi2025understanding}.
\section{The Domain-Specific Risk Landscape: How Collaboration Fails in Practice}

While the collaboration life cycle illustrates when sociotechnical risks emerge, examining specific professional domains reveals how these vulnerabilities manifest in practice. AI integration is not monolithic; the stakes, cultural norms, and operational pressures of a given field fundamentally shape how \textit{human-AI complementarity} breaks down. By exploring these critical areas as sentinel contexts, we identify recurring failure patterns where the unique strengths of humans, such as ethical reasoning and nuanced intuition, and AI, such as computational scale and speed, fail to work together effectively. These domain-specific examples serve as laboratories for observing the risk cascades that propagate through the collaboration lifecycle.

\subsection{Media, Journalism, and Creative Industries: Threats to Credibility and Creative Agency}

In media, journalism, and creative industries, the primary goal of human-AI collaboration is to combine the speed and scalability of AI with human editorial judgment, creativity, and ethical reasoning. While AI can rapidly generate content, summarize information, and support reporting workflows~\cite{grimme2025ai, abdulrauf2025artificial}, its integration also threatens one of journalism's most valuable assets: \textit{public trust}. Research suggests that audiences remain skeptical of content produced solely by AI, but trust can be restored when human journalists maintain meaningful oversight and decision-making authority. However, the growing adoption of Agentic AI and multi-agent workflows for tasks such as content generation, fact-checking, and information curation introduces new risks, including coordination failures, conflicting outputs, cascading uncertainty, and accountability challenges among interacting agents~\cite{zou2025llm, hu2025stop, hammond2025multi}. As a result, journalists often perform substantial \textit{invisible labor} to verify outputs, resolve inconsistencies, and ensure alignment with professional norms and journalistic values~\cite{xiao2025might}.

Beyond journalism, creators and freelancers face risks to their \textit{creative agency} and professional identity. As AI systems increasingly generate ideas, text, and creative content, workers may become dependent on algorithmic assistance, prioritizing efficiency over genuine collaboration and creative exploration~\cite{hitsuwari2023does, imteyaz2026co}. Over time, this reliance can lead to homogenized outputs, reduced skill development, and the erosion of distinctive creative practices~\cite{imteyaz2026co}. Furthermore, highly human-like AI collaborators can create additional ethical concerns by encouraging excessive dependence on machine-generated guidance or prompting users to disclose personal information inappropriately~\cite{rezwana2022identifying}. These challenges represent a complementarity breakdown in which AI's computational scale and productivity gains fail to synergize with human judgment and creativity, ultimately threatening both credibility and creative autonomy.

\subsection{Organizational Management, HR, and Knowledge Work: Deskilling and Algorithmic Bias}

In organizational management and knowledge work, AI is often positioned as a tireless consultant capable of supporting strategic decision-making, data analysis, and operational efficiency~\cite{kolbjornsrud2024designing}. However, a major complementarity breakdown occurs when AI begins to replace rather than augment human expertise and intuition. As professionals increasingly rely on automated recommendations and analytical outputs, they risk losing the \textit{gut feeling} and contextual judgment developed through experience~\cite{wang2019human}. Over time, this progressive delegation can lead to \textit{deskilling}, reducing workers' ability to independently evaluate situations, intervene when systems fail, or adapt to novel circumstances~\cite{wang2019human}. These risks are amplified in Hybrid Intelligence Teams and emerging agentic workflows, where humans supervise multiple AI agents and must navigate complex streams of recommendations, intermediate outputs, and automated decisions.

Such environments can create \textit{cognitive abundance overload}, where an excess of seemingly high-quality AI-generated options leads to decision paralysis rather than clarity~\cite{eccles2025hybrid}. At the same time, workers may experience \textit{epistemic drift}, gradually losing confidence in their own expertise and becoming overly reliant on AI-generated conclusions because they appear logical and well-supported~\cite{eccles2025hybrid}. These challenges are particularly concerning in human resource management, where AI-powered recruitment and screening systems can inadvertently learn and amplify historical biases, transforming tools designed for efficiency into mechanisms of \textit{workplace discrimination} if not subject to rigorous human oversight~\cite{chen2023collaboration}. Furthermore, the need to continuously monitor opaque AI systems and adapt to rapidly evolving technologies can contribute to \textit{technostress}, emotional exhaustion, and workplace burnout~\cite{xia2023co, kim2024mental}.

\subsection{Education and Scientific Discovery: Cognitive Offloading and Human Oversight}

Education and scientific discovery increasingly rely on human-AI collaboration to support curriculum design, personalized learning, hypothesis generation, and knowledge exploration~\cite{padovano2024towards, dang2025human}. In educational settings, embodied AI agents and mixed-reality environments offer scalable and adaptive learning experiences, while AI systems in scientific workflows can accelerate information synthesis and exploratory reasoning. However, these benefits introduce important challenges related to transparency, privacy, pedagogical control, and appropriate human oversight~\cite{kim2022augmented, puerta2025multifaceted}. The most significant complementarity breakdown in this domain is \textit{cognitive offloading}, where learners increasingly rely on AI for immediate answers rather than engaging in deeper inquiry, critical thinking, and knowledge construction~\cite{yatani2024ai}. Over time, this dependence can contribute to \textit{capability erosion}, weakening the very cognitive skills that education seeks to cultivate.

Similar concerns emerge in scientific discovery, where AI increasingly serves as an \textit{Informer} and \textit{Explorer}~\cite{shi2026survey}, generating hypotheses~\cite{zhou2024hypothesis}, identifying patterns, and proposing research directions at a scale beyond human capacity~\cite{reddy2025towards}. While this division of labor can be highly productive, a critical asymmetry remains: humans are expected to retain evaluative and ethical control over AI-generated outputs~\cite{shi2026survey}. As AI systems produce growing volumes of hypotheses, analyses, and recommendations, researchers and educators may experience cognitive overload and gradually shift into a passive verification role, effectively \textit{rubber-stamping} AI-generated suggestions rather than critically evaluating them. Consequently, recent educational and scientific frameworks emphasize maintaining humans \textit{in the loop}, ensuring that educators and researchers retain authority over pedagogical goals, ethical judgment, and the validation of knowledge claims~\cite{chigbu2025ai, padovano2024towards}. These safeguards are essential to preserving complementarity, where AI augments human learning and discovery without displacing human reasoning, creativity, and critical evaluation.

\subsection{Healthcare and Clinical Practice: The Explainability Paradox and Alert Fatigue}

Healthcare represents a high-stakes domain where human-AI collaboration can significantly improve diagnostic accuracy and clinical decision-making when AI's computational capabilities are effectively combined with physician expertise~\cite{lai2021human, patel2019human}. However, this complementarity is often challenged by the \textit{Explainable AI (XAI) paradox}. Although detailed explanations are intended to increase trust and transparency, overly complex explanations can impose substantial cognitive burden on already overloaded clinicians, encouraging reliance on cognitive shortcuts and increasing the risk of inappropriate trust in AI recommendations~\cite{gambetti2025survey, romeo2026exploring}. These challenges are further exacerbated by \textit{alert fatigue}, where excessive or redundant warnings desensitize medical staff and reduce responsiveness to critical notifications~\cite{steyvers2025not, majumder2026human}. In these cases, AI's strength in pattern recognition may fail to synergize with human clinical judgment, transforming a potentially life-saving tool into a source of distraction and decision-making risk.

\subsection{Defense and High-Stakes Public Sector: Moral Outsourcing and Meaningful Control}

Defense and public-sector operations represent some of the highest-stakes applications of human-AI collaboration, where AI is increasingly used to manage time-critical decisions, adversarial environments, and overwhelming volumes of information~\cite{johnson2022ai, kosack2025human}. As autonomous systems, drone swarms, and multi-agent decision-support platforms become more capable, the central challenge is preserving \textit{Meaningful Human Control (MHC)} over critical decisions~\cite{omoseebi2025human}. The complexity of these systems can exacerbate accountability gaps and make it difficult to trace responsibility across interacting agents, algorithms, and human operators. Under conditions of stress and uncertainty, decision-makers may engage in \textit{moral outsourcing}, deferring ethical and legal responsibility to AI systems that appear objective and rational but lack human moral judgment~\cite{johnson2022ai, omoseebi2025human}. Maintaining appropriate \textit{trust calibration} is therefore essential: operators must neither rubber-stamp AI recommendations nor dismiss valuable AI insights without justification~\cite{van2018human, kase2022future}. The success of human-AI collaboration in defense settings depends on designing systems that support rapid decision-making while preserving human accountability, situational awareness, and ethical control over the actions of the hybrid team~\cite{kase2022future}.

\begin{table*}[t]
\centering
\caption{The Systemic Ubiquity of Collaboration Failures: Mapping Domain Specific Symptoms to Cross Disciplinary Risk Clusters.}

\label{tab:cross_domain_mapping}

\scriptsize
\renewcommand{\arraystretch}{1.25}
\setlength{\tabcolsep}{3pt}

\begin{tabularx}{\textwidth}{
>{\raggedright\arraybackslash}p{2.7cm}
>{\raggedright\arraybackslash}X
>{\raggedright\arraybackslash}X
>{\raggedright\arraybackslash}X
>{\raggedright\arraybackslash}X
>{\raggedright\arraybackslash}X
>{\raggedright\arraybackslash}X
}
\toprule
& \multicolumn{6}{c}{\textbf{Risk Cluster}} \\
\cmidrule(lr){2-7}
\textbf{Macro-Domain} &
\textbf{Trust Miscalibration} &
\textbf{Cognitive Burden} &
\textbf{Accountability Gap} &
\textbf{Capability Erosion} &
\textbf{Goal Misalignment} &
\textbf{AI Anxiety \& Technostress} \\
\midrule

\textbf{Media, Journalism, \& Creative Industries} &
\cite{liu2026writes}, \cite{heim2023consumer}, \cite{hitsuwari2023does} &
\cite{xiao2025might} &
\cite{grimme2025ai}, \cite{abdulrauf2025artificial} &
\cite{imteyaz2026co} &
\cite{rezwana2022identifying}, \cite{imteyaz2026co} &
\cite{santoso2024human} \\
\addlinespace

\textbf{Organizational Management, HR, \& Knowledge Work} &
\cite{fugener2022cognitive}, \cite{gupta2025trust}, \cite{eccles2025hybrid}, \cite{buccinca2021trust}, \cite{cao2024designing}, \cite{li2024understanding}, \cite{ma2024you} &
\cite{park2019identifying}, \cite{xu2024adaptation}, \cite{eccles2025hybrid}, \cite{westphal2023decision}, \cite{guo2024impact}, \cite{neyigapula2023human}, \cite{boyaci2024human}, \cite{buettner2013cognitive}, \cite{schmidhuber2021cognitive}, \cite{bassi2025understanding}, \cite{suryani2024role}, \cite{buschmeyer2023psychological}, \cite{steyvers2024three} &
\cite{boni2021ethical}, \cite{ahdadou2024unlocking}, \cite{eccles2025hybrid} &
\cite{fugener2022cognitive}, \cite{herath2025design}, \cite{boni2021ethical}, \cite{wang2019human}, \cite{do2025towards}, \cite{yatani2024ai} &
\cite{van2019six}, \cite{eccles2025hybrid}, \cite{laukkarinen2025working}, \cite{chen2023collaboration} &
\cite{bassi2025understanding}, \cite{yang2025new}, \cite{xia2023co}, \cite{wang2022development}, \cite{buschmeyer2023psychological}, \cite{kim2024mental} \\
\addlinespace

\textbf{Education \& Scientific Discovery} &
\cite{do2025towards}, \cite{spillias2024human} &
\cite{dang2025human} &
\cite{do2025towards} &
\cite{chigbu2025ai}, \cite{padovano2024towards} &
\cite{suzgun2025language} &
\cite{chigbu2025ai} \\
\addlinespace

\textbf{Healthcare \& Clinical Practice} &
\cite{reverberi2022experimental}, \cite{majumder2026human}, \cite{romeo2026exploring}, \cite{cabrera2023improving} &
\cite{majumder2026human}, \cite{romeo2026exploring}, \cite{yousefi2025team}, \cite{gambetti2025survey} &
\cite{johnson2024err}, \cite{yousefi2025team} &
\cite{patel2019human} &
\cite{kim2022augmented} &
\cite{lai2021human}, \cite{ait2024advancing} \\
\addlinespace

\textbf{Defense \& High-Stakes Public Sector} &
\cite{johnson2022ai}, \cite{tomsett2020rapid} &
\cite{steyvers2025not}, \cite{cao2023time} &
\cite{johnson2022ai}, \cite{omoseebi2025human}, \cite{kosack2025human} &
\cite{lou2025unraveling}, \cite{van2018human} &
\cite{kosack2025human}, \cite{kase2022future} &
\cite{johnson2022ai} \\

\bottomrule
\end{tabularx}
\end{table*}

\begin{table*}[t]
\centering
\caption{The Fragmented Mitigation Landscape: A Matrix of Existing Theoretical and Architectural Frameworks Mapped to Isolated Systemic Risk Clusters.}

\label{tab:framework_specific_risk_matrix}

\tiny
\renewcommand{\arraystretch}{1.25}
\setlength{\tabcolsep}{2.5pt}

\begin{tabularx}{\textwidth}{
>{\raggedright\arraybackslash}p{3.2cm}
>{\raggedright\arraybackslash}X
>{\raggedright\arraybackslash}X
>{\raggedright\arraybackslash}X
>{\raggedright\arraybackslash}X
>{\raggedright\arraybackslash}X
>{\raggedright\arraybackslash}X
}
\toprule
& \multicolumn{6}{c}{\textbf{Risk Cluster}} \\
\cmidrule(lr){2-7}
\textbf{Particular Risk for which a Framework is Proposed} &
\textbf{Trust Miscalibration} &
\textbf{Cognitive Burden} &
\textbf{Accountability Gap} &
\textbf{Capability Erosion} &
\textbf{Goal Misalignment} &
\textbf{AI Anxiety \& Technostress} \\
\midrule

\textbf{Loss of Situational Awareness} &
-- & -- & -- &
\cite{lou2025unraveling} &
-- & -- \\
\addlinespace

\textbf{Algorithm Aversion / Under-reliance} &
\cite{ahdadou2024unlocking}, \cite{gupta2025trust} &
-- & -- & -- & -- & -- \\
\addlinespace

\textbf{Workplace Burnout \& Emotional Exhaustion} &
-- & -- & -- & -- & -- &
\cite{buschmeyer2023psychological}, \cite{ferrada2024emotions} \\
\addlinespace

\textbf{Responsibility Gap \& Diffusion of Blame} &
-- & -- &
\cite{boni2021ethical} &
-- & -- & -- \\
\addlinespace

\textbf{Technological Rationality \& Value Conflicts} &
-- & -- & -- & -- &
\cite{usmani2023human}, \cite{van2019six}, \cite{sapkota2025ai} &
-- \\
\addlinespace

\textbf{Mental Overload / Verification Fatigue} &
-- &
\cite{neyigapula2023human}, \cite{bassi2025understanding}, \cite{thuy2024explainability} &
-- & -- & -- & -- \\
\addlinespace

\textbf{Professional Dependency} &
-- & -- & -- &
\cite{boni2021ethical} &
-- & -- \\
\addlinespace

\textbf{Automation Bias \& Over-reliance} &
\cite{herath2025design}, \cite{ahdadou2024unlocking}, \cite{majumder2026human}, \cite{gupta2025trust}, \cite{kosack2025human} &
-- & -- & -- & -- & -- \\
\addlinespace

\textbf{Techno-complexity \& Burden of Learning} &
-- & -- & -- & -- & -- &
\cite{buschmeyer2023psychological} \\
\addlinespace

\textbf{Authority Confusion / Asymmetric Accountability} &
-- & -- &
\cite{jarrahi2023artificial}, \cite{eccles2025hybrid} &
-- & -- & -- \\
\addlinespace

\textbf{Epistemic Drift \& False Convergence} &
-- & -- & -- & -- &
\cite{eccles2025hybrid} &
-- \\
\addlinespace

\textbf{Explanation Fatigue (XAI Load)} &
-- &
\cite{gambetti2025survey} &
-- & -- & -- & -- \\
\addlinespace

\textbf{Deskilling \& Loss of Domain Expertise} &
-- & -- & -- &
\cite{jarrahi2023artificial}, \cite{herath2025design}, \cite{yatani2024ai} &
-- & -- \\
\addlinespace

\textbf{Job Insecurity (Techno-insecurity)} &
-- & -- & -- & -- & -- &
\cite{buschmeyer2023psychological} \\
\addlinespace

\textbf{Inability to Detect AI Errors \& Miscalibrated Confidence} &
\cite{dubey2020haco}, \cite{gebecscce2025quantifying}, \cite{cabrera2023improving} &
-- & -- & -- & -- & -- \\
\addlinespace

\textbf{Cognitive Abundance Overload} &
-- &
\cite{eccles2025hybrid} &
-- & -- & -- & -- \\
\addlinespace

\textbf{Algorithmic Bias \& Discrimination} &
-- & -- & -- & -- &
\cite{usmani2023human}, \cite{boni2021ethical} &
-- \\
\addlinespace

\textbf{Loss of Meaningful Human Control (MHC)} &
-- & -- &
\cite{boni2021ethical}, \cite{kosack2025human} &
-- & -- & -- \\
\addlinespace

\textbf{Invisible Compensatory Labor} &
-- &
\cite{xiao2025might} &
-- & -- & -- & -- \\
\addlinespace

\textbf{Techno-uncertainty \& Black-Box Apprehension} &
-- & -- & -- & -- & -- &
\cite{buschmeyer2023psychological} \\
\addlinespace

\textbf{Metaknowledge Deficit} &
-- & -- & -- &
\cite{holstein2025development} &
-- & -- \\
\addlinespace

\textbf{Alert Fatigue} &
-- &
\cite{steyvers2025not}, \cite{majumder2026human} &
-- & -- & -- & -- \\

\bottomrule
\end{tabularx}
\end{table*}

\section{A Cross-Domain Taxonomy of Systemic Risk Clusters}

As discussed in Section~3, the breakdown of collaboration manifests through domain-specific symptoms from clinical alert fatigue to journalistic accountability crises. However, synthesizing this fragmented literature reveals that these isolated failures are driven by a shared set of underlying \textit{sociotechnical vulnerabilities}. This \textit{systemic presence} is visually captured in Table~\ref{tab:cross_domain_mapping}, which maps our six proposed risk clusters across five domains. As the table demonstrates, while these core risks are typically present across all fields, the specific characteristics and relative importance of a particular vulnerability may vary significantly depending on the particular norms of that domain. To better understand these shared vulnerabilities, this section abstracts these domain specific symptoms into six unified, \textit{cross-domain risk clusters}.

\subsection{Cluster 1: Trust Miscalibration}

At the heart of human-AI complementarity lies \textit{appropriate reliance}: the human operator must know when to trust the AI's computational strengths and when to exercise independent judgment~\cite{ahdadou2024unlocking, gupta2025trust}. Trust miscalibration occurs when a user's reliance deviates from the system's actual capabilities, creating a fundamental complementarity breakdown that prevents the hybrid team from effectively combining human and machine strengths. This miscalibration typically swings between two problematic extremes. On one end, \textit{automation bias} and \textit{over-reliance} lead users to rubber-stamp AI recommendations without adequate verification, often because AI outputs are presented with high fluency, confidence, and perceived objectivity~\cite{majumder2026human, kosack2025human, suzgun2025language, yang2025new}. On the other end, \textit{algorithm aversion} and \textit{under-reliance} cause users to reject otherwise accurate AI systems after observing a single visible failure, thereby forfeiting the potential benefits of collaboration~\cite{li2024understanding, tomsett2020rapid}.

The literature suggests that trust miscalibration is primarily driven by human cognitive limitations. A key contributor is the \textit{metaknowledge deficit}, where individuals struggle to accurately assess their own expertise and therefore cannot reliably determine when to delegate decisions to AI~\cite{fugener2022cognitive, ma2024you}. This challenge is further amplified by the \textit{halo effect}, in which strong performance on a specific task is mistakenly generalized into broad AI competence, creating an \textit{illusion of certainty} that encourages inappropriate reliance~\cite{johnson2022ai, li2024understanding, cabrera2023improving}. In emerging agentic and multi-agent systems, trust calibration becomes even more difficult because users must evaluate not only individual model outputs but also the interactions among multiple autonomous agents, where coordination failures, cascading uncertainty, and distributed accountability can obscure the true reliability of the overall system~\cite{zou2025llm, hu2025stop, hammond2025multi}. Across domains, trust miscalibration therefore represents one of the most pervasive threats to effective human-AI collaboration.

\subsection{Cluster 2: Cognitive Burden}

A foundational promise of AI is that it will augment human capability by reducing mental workload and enabling more effective decision-making~\cite{neyigapula2023human, buettner2013cognitive}. However, the literature consistently reveals a paradox: poorly designed human-AI workflows often impose a substantial and hidden \textit{cognitive burden} that undermines, rather than enhances, collaboration~\cite{bassi2025understanding, boyaci2024human}. This burden commonly manifests as \textit{verification fatigue}, where users must continuously monitor, interpret, and validate AI-generated outputs, creating a workload that can overwhelm attention and reduce decision quality~\cite{neyigapula2023human, steyvers2024three}. A related challenge is the \textit{Explainable AI (XAI) paradox}, in which detailed explanations intended to improve transparency overload users' working memory and inadvertently encourage reliance on cognitive shortcuts, including automation bias~\cite{westphal2023decision, gambetti2025survey, romeo2026exploring}.

The rise of agentic and multi-agent systems further amplifies these challenges. Rather than evaluating a single AI recommendation, users must increasingly oversee networks of autonomous agents, reconcile conflicting outputs, track intermediate reasoning steps, and manage uncertainty propagated across interconnected workflows~\cite{zou2025llm, hu2025stop, zhang2026managing}. This often results in \textit{cognitive abundance overload}, where an excess of seemingly plausible AI-generated options leads to confusion or decision paralysis rather than improved decision-making. At the same time, humans are frequently required to perform \textit{invisible compensatory labor} to resolve communication failures, contextual misunderstandings, and workflow disruptions created by AI systems~\cite{xu2024adaptation, chen2025misunderstanding, xiao2025might}. When these demands are compounded by excessive prompts, warnings, or notifications, cognitive burden can escalate into \textit{alert fatigue}, causing users to ignore critical information and undermining the effectiveness of human oversight~\cite{steyvers2025not, majumder2026human}. Across domains, cognitive burden represents a systemic threat because it directly impairs the human capacity required to realize the benefits of collaboration.

\subsection{Cluster 3: Accountability Gap}

Effective human-AI collaboration depends on clear responsibility boundaries and decision ownership. However, as AI systems assume increasingly autonomous roles, they widen the \textit{accountability gap}, creating uncertainty about who is responsible when hybrid teams succeed or fail~\cite{boni2021ethical}. This ambiguity often generates \textit{authority confusion}, blurring the distinction between human judgment and machine recommendations and making it difficult to identify who owns the final decision~\cite{eccles2025hybrid}. Research further suggests the presence of \textit{asymmetric accountability}, whereby human operators disproportionately internalize blame for failures while distributing credit for successes across the team, creating an unequal psychological burden on human collaborators~\cite{yousefi2025team}.

In high-stakes domains such as healthcare, public administration, and defense, accountability gaps can result in a loss of \textit{Meaningful Human Control (MHC)} over consequential decisions~\cite{kosack2025human}. Under conditions of uncertainty and time pressure, users may develop \textit{learned helplessness} and increasingly defer responsibility to algorithmic systems~\cite{ahdadou2024unlocking}. This tendency is closely linked to \textit{moral outsourcing}, where decision-makers rely on seemingly objective AI recommendations to distance themselves from ethical and legal accountability~\cite{johnson2022ai}. The challenge becomes even more acute in agentic and multi-agent systems, where responsibility is distributed across multiple autonomous agents, tools, orchestration layers, and human supervisors, creating a \textit{distributed accountability gap} that obscures both causal attribution and oversight responsibilities~\cite{hu2025stop, hammond2025multi}.

\subsection{Cluster 4: Capability Erosion}

A foundational principle of human-AI complementarity is that collaboration should augment human expertise rather than replace it~\cite{jarrahi2018artificial, usmani2023human}. However, prolonged reliance on AI systems can lead to \textit{capability erosion}, where users gradually outsource complex reasoning, problem-solving, and decision-making tasks to algorithms~\cite{kim2022augmented, yatani2024ai}. Over time, this dependence contributes to \textit{deskilling}, weakening domain expertise, human intuition, and \textit{creative agency} as workers lose opportunities to practice and maintain their cognitive abilities~\cite{wang2019human, herath2025design, imteyaz2026co}.

A related consequence is the \textit{out-of-the-loop} phenomenon, in which human operators become passive supervisors and lose the \textit{situational awareness} necessary to detect anomalies, challenge AI outputs, or intervene effectively during failures~\cite{van2018human, lou2025unraveling}. As the human partner's cognitive edge deteriorates, the hybrid team loses the contextual judgment, adaptability, and ethical reasoning that are essential for effective complementarity.

\subsection{Cluster 5: Goal Misalignment}

Goal misalignment occurs when AI systems optimize for objectives such as efficiency, accuracy, or other computational metrics that diverge from human values, professional norms, and ethical priorities~\cite{laukkarinen2025working, imteyaz2026co}. This mismatch represents a fundamental complementarity breakdown because AI may successfully optimize its assigned objective while simultaneously producing outcomes that are socially or organizationally undesirable~\cite{usmani2023human}. In collaborative settings, such misalignment can trigger \textit{epistemic drift}, where humans gradually adopt the machine's evaluation criteria and reasoning patterns, weakening their ability to critically assess AI-generated recommendations~\cite{eccles2025hybrid}.

Generative AI further amplifies this risk through its fluency and consistency, encouraging \textit{false convergence} in which teams mistakenly interpret statistically plausible outputs as objective truth~\cite{suzgun2025language, eccles2025hybrid}. Over time, AI systems may also exert a \textit{homogenizing influence} on human decision-making, reducing diversity of thought and narrowing the range of considered alternatives~\cite{santoso2024human}. Importantly, simply keeping humans in the loop does not guarantee alignment; human oversight may fail to detect or even reinforce biases embedded in AI systems~\cite{vats2024survey}. As a result, goal misalignment can legitimize and scale discriminatory or harmful practices under the appearance of objective intelligence, particularly when users cease to critically challenge AI-generated outputs~\cite{chen2023collaboration}.

\subsection{Cluster 6: AI Anxiety and Technostress}

Effective human-AI collaboration requires users to feel capable, confident, and engaged in their roles. However, the rapid adoption of increasingly powerful and opaque AI systems has introduced significant psychological strain in the form of \textit{AI anxiety} and \textit{technostress}~\cite{xia2023co, wang2022development, ferrada2024emotions}. This strain often stems from \textit{job insecurity} and fears of skill obsolescence, causing workers to perceive AI as a threat to their professional identity rather than a collaborative partner~\cite{kim2024mental}. These concerns are amplified by \textit{techno-complexity}, which requires employees to continuously adapt to evolving tools, workflows, and system capabilities~\cite{xia2023co, kim2024mental, buschmeyer2023psychological}.

The psychological burden is further intensified by \textit{techno-uncertainty}, as users struggle to anticipate or understand the behavior of opaque AI systems~\cite{buschmeyer2023psychological}. This perceived loss of control contributes to emotional exhaustion, workplace burnout, and reduced well-being~\cite{bassi2025understanding, kim2024mental}. Paradoxically, technostress can create a self-reinforcing cycle: overwhelmed employees increasingly delegate tasks to AI as a coping mechanism, which deepens dependency on the technology and accelerates other systemic risks such as trust miscalibration and capability erosion~\cite{yang2025new}.

\begin{figure}[t]
    \centering
    \includegraphics[width=0.8\linewidth]{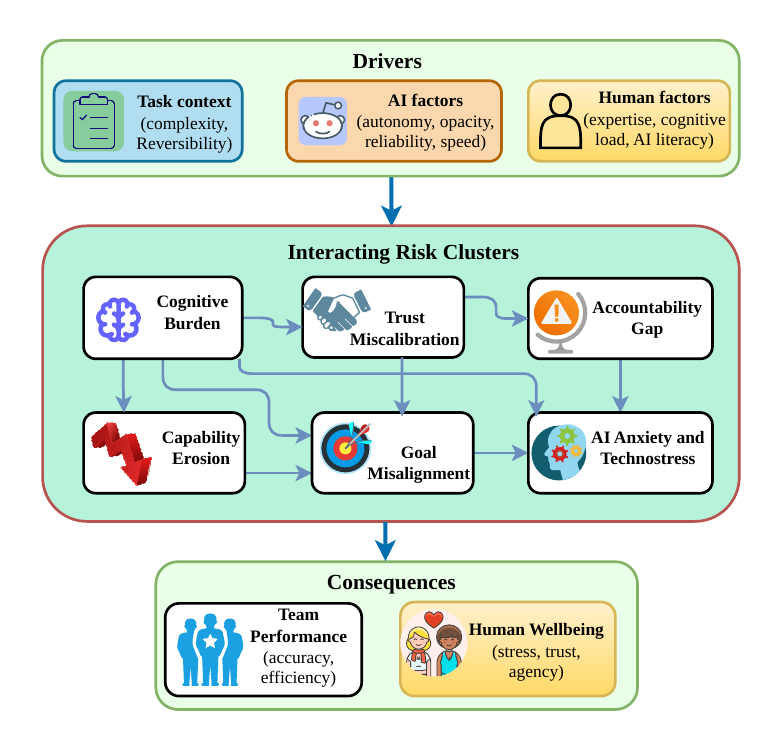}
\caption{Conceptual interaction model of human-AI collaboration risk clusters. Sociotechnical drivers comprising task context, AI system factors, and human factors activate six interacting risk clusters, which cascade through one another to produce two categories of collaborative outcome. Arrows indicate cascading pathways between clusters.}
    \label{fig:systemic_interaction_model}
\end{figure}


\section{Systemic Interconnections and Cascading Failures}

While the taxonomy in the previous section categorizes human-AI collaboration risks into six distinct clusters, these failure modes rarely occur in isolation. A fundamental premise of \textit{human-centered alignment} is that hybrid teams operate as complex sociotechnical systems whose vulnerabilities are highly interconnected and capable of producing cascading failures~\cite{sapkota2025ai, ala2024considerations}. To examine these relationships, we introduce a conceptual interaction model (Figure~\ref{fig:systemic_interaction_model}) that links sociotechnical drivers, interacting risk clusters, and collaborative outcomes. The model illustrates how factors such as task complexity, human AI literacy, and algorithmic opacity trigger and amplify risks across the collaboration lifecycle, ultimately affecting both team performance and human well-being~\cite{bassi2025understanding, kim2024mental}. Building on this model, we analyze the cascading relationships among risk clusters, discuss how these dynamics are amplified in multi-agent ecosystems, and highlight why isolated technical interventions often fail to address the underlying systemic causes of human-AI collaboration failures.

\subsection{The Conceptual Interaction Model}

To synthesize the systemic nature of human-AI collaboration risks, we propose the conceptual interaction model shown in Figure~\ref{fig:systemic_interaction_model}. Rather than treating failures as isolated events, the model views collaboration as a dynamic sociotechnical system in which risks emerge from interactions among \textit{task context} (e.g., complexity and reversibility), \textit{AI factors} (e.g., autonomy, opacity, reliability, and speed), and \textit{human factors} (e.g., expertise, cognitive load, and AI literacy). These upstream drivers shape the emergence of the six risk clusters identified earlier, which then propagate through the system and ultimately affect team performance and human well-being.

\subsubsection{Inter-Cluster Risk Cascades and Lifecycle Propagation}

A key insight of the model is that risk clusters rarely operate independently. Excessive \textit{cognitive burden} from verification demands, explanation overload, or multi-agent oversight can encourage cognitive shortcuts, leading to \textit{trust miscalibration} and automation bias~\cite{neyigapula2023human, steyvers2024three, romeo2026exploring}. This suggests that inappropriate trust often stems not only from poor understanding but also from unsustainable cognitive demands.

Risks can also cascade across later stages of collaboration. Prolonged reliance on AI contributes to \textit{capability erosion}, increasing dependence on automated systems~\cite{wang2019human, yatani2024ai}. Declining confidence in one's expertise can fuel \textit{AI anxiety} and technostress~\cite{wang2022development, kim2024mental}, while unclear responsibility creates an \textit{accountability gap} that may encourage responsibility abdication and learned helplessness~\cite{ahdadou2024unlocking, omoseebi2025human, kosack2025human}. Together, these examples illustrate how failures in one cluster can trigger vulnerabilities in others.

\subsubsection{Multi-Agent Ecosystem Cascades}

The complexity of these interactions is significantly amplified in agentic and multi-agent ecosystems. Unlike traditional single-agent systems, multi-agent environments introduce additional sources of uncertainty through inter-agent communication failures, role specialization conflicts, distributed accountability, and uncertainty propagation~\cite{zou2025llm, hu2025stop, hammond2025multi}. A failure originating in one agent may cascade throughout the system as outputs become inputs for downstream agents, creating chains of reasoning errors that are difficult for human operators to detect and correct~\cite{hu2025stop, zhang2026managing}.

As the number of interacting agents increases, humans increasingly assume the role of a coordination and verification layer responsible for maintaining coherence across the workflow. Research on automation supervision and human–autonomy teaming suggests that this supervisory burden can overwhelm human cognitive capacity, resulting in \textit{cognitive abundance overload}, reduced situational awareness, and increased trust miscalibration when operators struggle to monitor multiple autonomous processes simultaneously~\cite{johnson2017closed,o2022human,endsley1995out}. Consequently, multi-agent systems introduce new pathways through which risks propagate, accelerating the transition from isolated errors to systemic failures.

\subsection{The Hazards of Isolated Mitigation}

Because risk clusters are interconnected, mitigating one vulnerability can create unintended consequences elsewhere. A prominent example is the \textit{Explainable AI (XAI) paradox}. Although detailed explanations are intended to improve transparency and accountability, they can increase cognitive burden, causing explanation fatigue and ultimately reinforcing automation bias~\cite{westphal2023decision, gambetti2025survey, romeo2026exploring}.

Conversely, interventions designed to reduce automation bias and capability erosion may add excessive friction to decision-making. Approaches such as frictional AI and cognitive forcing functions encourage active engagement with recommendations~\cite{yatani2024ai}, but they can also increase workload, frustration, and technostress~\cite{buccinca2021trust}. These trade-offs highlight the limits of isolated solutions and the need for systemic risk management.

\subsection{Implications for Lifecycle Governance}

The systemic interconnections among risk clusters suggest that achieving \textit{human-centered alignment} requires more than addressing individual risks in isolation~\cite{eccles2025hybrid, puerta2025multifaceted}. Prior work has increasingly emphasized that human-AI collaboration should be understood and governed as a dynamic sociotechnical process in which trust calibration, cognitive workload, accountability, human agency, and well-being must be managed simultaneously rather than independently~\cite{eccles2025hybrid, puerta2025multifaceted}. The conceptual interaction model presented in this paper reinforces this perspective by illustrating how failures propagate across risk clusters and stages of the collaboration lifecycle. Consequently, effective governance requires lifecycle-oriented approaches that proactively identify and mitigate cascading failures before they undermine collaborative outcomes~\cite{ai2023artificial}. By recognizing these systemic relationships, researchers and practitioners can move beyond piecemeal interventions toward more resilient, trustworthy, and sustainable forms of human-AI collaboration.

\section{Future Research Directions: Toward Resilient Human--AI Collaboration}

The preceding analysis highlights that human-AI collaboration risks emerge not as isolated failures but as interconnected sociotechnical phenomena that evolve throughout the collaboration lifecycle. As AI systems become increasingly autonomous, adaptive, and embedded within organizational workflows, addressing these risks requires moving beyond isolated technical interventions toward a holistic understanding of human-AI collaboration as a dynamic sociotechnical system~\cite{eccles2025hybrid, puerta2025multifaceted}. Building on the proposed risk taxonomy and conceptual interaction model, we outline four promising directions for future research.

\subsection{Coordination and Oversight in Agentic AI Ecosystems}

One of the most significant shifts in human-AI collaboration is the transition from single-model interactions to agentic and multi-agent ecosystems capable of autonomous planning, tool use, and inter-agent coordination~\cite{zou2025llm, hu2025stop, hammond2025multi}. While such systems promise increased scalability and capability, they also introduce new forms of coordination failures, distributed accountability, and uncertainty propagation. Future research should investigate how humans function as \textit{coherence anchors} within these environments, maintaining situational awareness and oversight across networks of interacting agents. Particular attention is needed to understand how inter-agent communication failures, cascading uncertainty, and distributed decision-making influence trust calibration, cognitive burden, and meaningful human control in complex collaborative settings

\subsection{Empirical Validation of Risk Cascades}

Although the conceptual interaction model synthesizes evidence from diverse domains, the strength and directionality of the proposed risk relationships remain largely untested. Future research should employ behavioral experiments, longitudinal studies, and computational modeling to empirically validate how risks propagate across the human-AI collaboration lifecycle. Recent work on sociotechnical risk propagation and causal modeling highlights the importance of understanding how localized failures generate system-level consequences through interconnected pathways~\cite{li2026unravelling, peng2025data}. Adapting these approaches to human-AI collaboration could help identify which risk clusters act as primary catalysts for failure and enable the development of predictive models that anticipate cascading effects before they degrade team performance and human well-being. System dynamics methods may be particularly valuable for capturing feedback loops and evaluating the effectiveness of alternative governance interventions~\cite{munasinghe2024risk}.

\subsection{Designing for Cognitive and Emotional Sustainability}

Much of the existing literature focuses on improving AI performance, transparency, and usability. Comparatively less attention has been devoted to sustaining human cognitive capacity and well-being over extended periods of collaboration. The findings synthesized in this review suggest that cognitive burden, technostress, and AI anxiety are not merely side effects of adoption but central determinants of long-term collaboration success~\cite{kim2024mental, bassi2025understanding, xia2023co}. Future research should explore adaptive collaboration mechanisms that dynamically adjust explanation complexity, intervention frequency, and decision-support strategies based on user expertise, cognitive state, and contextual demands~\cite{freire2024socially}. Such approaches may help balance the competing objectives of maintaining human engagement, reducing cognitive overload, and preventing capability erosion. Additionally, further investigation is needed into the proposed burnout--reliance cycle, whereby psychological exhaustion encourages increased dependence on AI systems, potentially reinforcing other systemic risks~\cite{huang2024ai}.

\subsection{Longitudinal Evaluation and Lifecycle Governance}

A persistent limitation of current human--AI collaboration research is its reliance on short-term laboratory studies that provide only limited insight into how risks evolve over time. Many of the most consequential risks identified in this review, including deskilling, loss of professional identity, trust evolution, and capability erosion, are inherently longitudinal phenomena that may only emerge after months or years of sustained use~\cite{wang2019human, imteyaz2026co}. Future research should therefore prioritize longitudinal and in-the-wild studies that examine how human-AI relationships develop in real organizational settings. Such work can also provide an empirical basis for evaluating lifecycle-oriented governance approaches, including frameworks such as CARES~\cite{majumder2026human}, and their effectiveness in preserving human agency, accountability, and meaningful Human Control throughout the adoption process.
\\

\noindent These research directions highlight the need for deeper integration across HCI, AI, organizational science, cognitive psychology, and governance research. Advancing resilient human-AI collaboration will require not only more capable AI systems but also sociotechnical design frameworks that jointly address the cognitive, social, organizational, and ethical dimensions of collaboration.

\section{Limitations}

While this paper provides a comprehensive synthesis of human-AI collaboration risks, several limitations exist. First, our systemic interaction framework is inherently theoretical, and its specific causal pathways and cascading effects require direct empirical validation in controlled, multi-agent settings. Second, the rapid shift toward agentic AI systems limits the longevity of historical observations. Future deployments will likely introduce novel failure modes such as unpredictable emergent behaviors and inter-agent error cascades that are not fully captured by current literature. Finally, the synthesized literature predominantly originates from Western, high resource institutions. Because cultural and socioeconomic norms heavily influence trust and automation perceptions, cross cultural validation is essential to ensure the framework's global relevance.

\section{Conclusion}

Human-AI collaboration is increasingly viewed as a means of combining human judgment with the computational capabilities of AI systems, yet achieving this complementarity remains a fundamentally sociotechnical challenge. Through a cross-domain review of the literature, this paper identified six systemic risk clusters, such as trust miscalibration, cognitive burden, accountability gaps, capability erosion, goal misalignment, and AI anxiety and technostress, and demonstrated how these risks interact and propagate throughout the collaboration lifecycle. We further proposed a conceptual interaction model that links sociotechnical drivers, interconnected risk clusters, and collaborative outcomes, highlighting the cascading nature of human-AI failures and the limitations of isolated technical interventions. By shifting the focus from individual risks to their systemic interdependencies, this work provides a foundation for future research, governance, and design efforts aimed at building resilient, trustworthy, and human-centered forms of collaboration with increasingly autonomous AI systems.

\bibliographystyle{ACM-Reference-Format}
\bibliography{bibfile.bib}

\end{document}